# Reconfigurable Room Temperature Exchange Bias through Néel Order Switching in van der Waals Heterostructures

Jicheng Wang[1], Shilei Ding[2]*, Bei Ding[3], Zhipeng Hou[3]*, Licong Peng[4], Yilan Jiang[5], Fengshan Zheng[1,5], Zhaochu Luo[6]*, Yu Ye[6], Jinbo Yang[6], Yanglong Hou[4,7]*, Rui Wu[1]*

1. Spin-X Institute, School of Physics and Optoelectronics, State Key Laboratory of Luminescent Materials and Devices, and Guangdong-Hong Kong-Macao Joint Laboratory of Optoelectronic and Magnetic Functional Materials, South China University of Technology, Guangzhou 511442, China
2. Department of Materials, ETH Zürich, 8093 Zürich, Switzerland
3. Guangdong Provincial Key Laboratory of Optical Information Materials and Technology, Institute for Advanced Materials, South China Academy of Advanced Optoelectronics, South China Normal University, Guangzhou 511442, China
4. School of Materials Science and Engineering, Peking University, Beijing 100871, P.R. China
5. Center for Electron Microscopy, South China University of Technology, Guangzhou 511442, China
6. State Key Laboratory for Mesoscopic Physics, School of Physics, Peking University, Beijing 100871, P.R. China
7. School of Materials, Shenzhen Campus of Sun Yat-Sen University, Shenzhen 518107, China

*Corresponding author: shilei.ding@mat.ethz.ch, houzp@m.scnu.edu.cn, zhaochu.luo@pku.edu.cn, hou@sysu.edu.cn, and ruiwu001@scut.edu.cn

**Abstract**

  Exchange bias effect plays a crucial role in modern magnetic memory technology. Recently, van der Waals magnetic materials have emerged and shown potential in spintronic devices at atomic scale. Owing to their tunable physical properties and the flexibility in fabrication, the van der Waals heterostructures offer more possibilities for investigating potential mechanisms of the exchange bias effect. However, due to low magnetic ordering temperatures for most van der Waals magnets, to establish exchange bias in van der Waals antiferromagnet/ferromagnet heterostructures at room temperature is challenging. In this study, we fabricate $(Fe_{0.56}Co_{0.44})_5GeTe_2$(FCGT)/$Fe_3GaTe_2$(FGaT) heterostructures with magnetic ordering temperatures of each component well above room temperature to achieve a room temperature exchange bias effect. It is found that the sign and magnitude of the exchange bias field can be efficiently controlled by manipulating the Néel order of FCGT with magnetic field. The manipulation of Néel order shows significant magnetic field dependence. A strong pre-set field induces a switch in the Néel order of FCGT, which aligns the interfacial magnetization at the FCGT/FGaT interface, leading to robust exchange bias, as revealed by both transport measurements and macro-spin model calculations. Our findings demonstrate the intrinsic manipulation and switchable of room-temperature exchange bias in all-van der Waals heterostructures and further promote the development of novel two-dimensional spintronic devices.

**Introduction**

The interfacial exchange coupling between ferromagnets (FM) and antiferromagnets (AFM), manifesting as unidirectional anisotropy, leads to a horizontal shift in magnetic hysteresis, a phenomenon known as the exchange bias (EB) effect. The EB phenomenon has great practical application in magnetic recording technologies. For example, in the magnetic tunnel junction (MTJ), the core component of magnetic random-access memory (MRAM) (*1, 2*), the EB effect is used to pin the magnetization of the reference layer, thereby achieving two different magnetoresistance states during the switching of magnetization of the free layer. The induction of EB in traditional and thin film materials usually require field cooling process, which is not conducive to modify.

The long-range magnetic order in van der Waals (vdW) materials, persisting down to the two-dimensional (2D) monolayer scale and up to room temperature, highlights their significant potential for next-generation spintronic devices (*3-6*). The weak interlayer exchange coupling, resulting from the vdW interaction (*7, 8*) makes this system an ideal platform for the manipulation of magnetic structures and the exploration of novel magnetic properties such as noncollinear spin structures (*9-12*), coherent and incoherent magnon excitations (*13-16*), and exchange interactions at interfaces (*17, 18*). The vdW materials enable the creation of abundant interfaces by stacking different materials to form heterostructures with their controllable thickness and atomically flat surfaces (*19, 20*). This facilitates precise control of heterostructure properties, providing an ideal platform to establish and modify the EB for the development of novel spintronic devices.

Recent work reported the discovery of EB in different vdW AFM/FM heterostructures, such as $CrCl_3$/$Fe_3GeTe_2$ (*21*), $CrPS_4$/(O-FGT)/$Fe_3GeTe_2$ (*22*), $FePS_3$/$Fe_3GaTe_2$ (*23*), $Fe_3GeTe_2$/CrOCl (*24*), O-$Fe_3GeTe_2$/$Fe_3GeTe_2$ (*25, 26*), $CrI_3$/$MnBi_2Te_4$ (*27*), and CrSBr/$Fe_3GaTe_2$ (*17*). However, the EB in most vdW heterostructures exhibits a weak EB field and extremely low blocking temperature ($T_B$) at which EB disappears, as the Néel temperature ($T_N$) of vdW AFM is typically low. Thus, great efforts have been made to alter the EB field and increase the blocking temperature, including methods like pressure engineering (*28*) and gate voltage application (*29, 30*). While these extrinsic approaches have advanced the study and modification of the EB, the field-cooling process down to the Néel temperature of the AFM remains necessary to establish EB in most vdW heterostructures. A novel approach to establish EB interaction was introduced by applying a pre-set field higher than the maximum field used for subsequent magnetic measurements (*17, 22, 31, 32*). This opens up a new strategy to efficiently generate and control the EB. Specifically, the establishment of EB by applying the pre-set field in single layer of vdW ferromagnet is attributed to the presence of local regions with higher coercivity, induced by pinning sources originating from crystal defects (*32, 33*). However, such systems typically suffer from strong training effects, and EB may vanish when the device size shrinks as the effect is related to the local defects in the samples. A further question arises as to whether a pre-set field can induce and modify the strong unidirectional interfacial exchange interactions in AFM/FM heterostructures, leading to the establishment of EB that is free from training effects and can be easily controlled by the pre-set field.

Here, we report the pre-set field induced robust EB in $(Fe_{0.56}Co_{0.44})_5GeTe_2$ (FCGT)/ $Fe_3GaTe_2$ (FGaT) heterostructures. By using FM and AFM materials with magnetic ordering temperatures both above the room temperature (RT), an EB effect up to RT ($T_B$ = 300 K) is established in the heterostructures. Both the sign and magnitude of the EB in this system can be efficiently controlled by manipulating the Néel order of FCGT with varying external pre-set field, allowing for a reconfigurable multi-state EB effect. Especially, a stable and training-free EB effect is achieved by applying a strong pre-set field beyond the critical field that could align the Néel order of FCGT, while a low pre-set field cannot align the uncompensated moment at the interface of FCGT, preventing the establishment of stable unidirectional exchange coupling in the FCGT/FGaT heterostructures. A macrospin model also confirms the switch of Néel order in FCGT, which explain the observed EB effect.

**Results**

**Fabrication and characterization of FCGT/FGaT heterostructures**

FCGT exhibits itinerant A-type antiferromagnetic ordering, and its easy axis of magnetization is along the crystallographic *c*-axis. $T_N$ of single crystal samples reaches approximately 335 K (*34, 35*). FGaT, with perpendicular magnetic anisotropy (PMA) and high Curie temperature ($T_C$) of 340 ~ 380 K (*3, 10, 36*), is employed to fabricate FCGT/FGaT heterostructures for the generation of a perpendicular EB effect. The FCGT crystals in this study were prepared using the CVT method, and FGaT crystals were prepared by self-fluxing method (See Materials and Methods section). The crystal structure and quality of both were verified by the X-ray diffraction and scanning electron microscope (Fig. S1), and the results indicate the good crystallization of our FCGT and FGaT samples.

To investigate the EB effect, we fabricate FCGT/FGaT heterostructures (as illustrated in Fig. 1A) and further create Hall bar device for the subsequence transport measurement. Fig. 1B illustrates the fabrication process of FCGT/FGaT Hall bar device (Device#1). FGaT flakes were mechanically exfoliated onto $SiO_2$/Si substrates using Scotch tape, followed by polycarbonate (PC)/polydimethylsiloxane (PDMS)-mediated transfer of FCGT flakes onto selected FGaT flakes in glove box under nitrogen atmosphere to prevent surface oxidation. The heterostructure was subsequently patterned into a Hall bar device via maskless UV lithography, plasma etching, and magnetron sputtering. Fig. S2A shows the optical image of Device#1. The interfacial structure was studied through cross-sectional scanning transmission electron microscopy (STEM) and energy-dispersive X-ray spectroscopy (EDS) mapping. Atomic-resolution high angle annular dark field scanning transmission electron microscopy (HAADF-STEM) imaging (Fig. S3A and B) demonstrates a flat interface with well-preserved crystallinity in both FCGT and FGaT layers. An amorphous oxide layer of 5 nm can be seen at the interface due to the inevitable exposure to air during all the measurement (Fig. S3C). EDS elemental mappings (Fig. S3D) confirm distinct spatial distributions of constituent elements within FCGT and FGaT layers. The EDS result also shows that the amorphous layer has a clear boundary in the middle, the upper layer contains Fe, Ge, O elements, and the lower layer contains Fe, Ga, O elements. Electron energy-loss spectroscopy (EELS) is performed to check possible phase of iron oxide (Fig. S3E ~ H). Fig. S3G illustrates the Fe-L-edge EELS spectra of three layers. The position of $L_3$ and $L_2$ peaks confirms the presence of iron element. Among them, the $L_3/L_2$ ratio of iron in the amorphous layer is calculated to be 4.5, and its $L_3$ peak is located at a relatively higher value of 709.7 eV, which is consistent with these characteristic parameters of α-$Fe_2O_3$ in the literature (*37*), indicating that the valence state of iron in the amorphous layer is a stable trivalent (+3). In contrast, the $L_3$ peak of iron in the FCGT and FGaT layers is split, and the relatively lower peak position represents the existence of divalent iron, proving the mixed valence state of iron in both, similar to $Fe_3GeTe_2$ (*38*). Therefore, the iron in FGaT and FCGT is easily oxidized, especially for FGaT (Fig. S3D). Even if the transfer process is carried out in a glove box, the appearance of the oxide layer cannot be completely avoided, which is consistent with the situation in many other FGaT-based heterostructures (*24, 39-41*).

The magnetic properties of FCGT and FGaT crystals were first characterized. Fig. 1C shows the temperature dependence of the magnetization of a FCGT single crystal measured under an applied magnetic field of 0.1 T. The measurements included zero field-cooling (ZFC) and field-cooling (FC) protocols, with the applied magnetic fields along the in-plane (IP; $H//ab$ plane) and out-of-plane (OOP; $H//c$ axis) directions, respectively. We can observe that both the ZFC and FC measurement exhibit a broad peak when the applied magnetic field is along these two directions, indicating a distribution of superparamagnetic $T_B$' centered at 275 K (*42*), i.e. the temperature at which the antiferromagnetic domains become unstable against the thermal fluctuation. The $T_N$ is identified to be in the range of 340 ~ 370 K, as characterized by the temperature-dependent anomalous Hall effect (AHE) measurements (Fig. S4A and B). Below 12 K, the magnetization of the ZFC curve drops sharply compared to the FC curve, indicating the existence of a spin freezing state in FCGT at temperatures below 12 K.

The magnetic hysteresis loops (*M-H* curves) of the FCGT crystal at 5 K for OOP configuration is shown in Fig. S5A. The spin-flop and spin-flip transitions can be observed, which is a typical characteristic of an A-type antiferromagnet, where the easy axis is parallel to the *c* axis. The *M-H* curve shows a large hysteresis at low temperatures because the spin-flop transition can be pinned by the surface or defect position, which is unavoidable in doped materials (*34*). In particular, during the initial magnetization process, as the magnetic field increases, the magnetization shows a linear relationship with the magnetic field and saturates at approximately 80 mT, exhibiting ferromagnetic characteristics. This behavior may originate from uncompensated magnetic moments due to the magnetic defects (*32, 33*). As the magnetic field continues to increase, the magnetization shows little changes, corresponding to the antiferromagnetic ground state. When the magnetic field reaches 2.3 T, a sharp nonlinear increase is observed, indicating a field-induced spin-flop transition, where the magnetic moments suddenly deviate from the *c* axis direction and align in *ab* plane. With further increasing the magnetic field, the magnetic moments gradually tilt towards the magnetic field direction, and the magnetization increases linearly with *H*, finally saturating at spin-flip field ($H_{flip}$) of about 5.8 T. Notably, $H_{flip}$ decreases gradually with increasing temperature (Fig. S5B).

The magnetism properties of the FGaT flake are shown in Fig. 1D. The ZFC and FC measurement under a 0.1 T magnetic field reveal typical ferromagnetic characteristics, with a $T_C$ as high as ~366 K as show in Fig. 1D. To further investigate the temperature-dependent magnetotransport properties, we fabricated a six-electrode Hall device using an exfoliated FGaT flake. (The optical image is shown in Fig. S2B.)

The AHE loops were obtained when an external magnetic field $H$ was applied parallel to the $c$ axis of the flake, as shown in Fig. S6A. When the temperature is below 300 K, the field-dependent AHE loops ($R_{xy}$-$H$) exhibit rectangular shapes, indicating that FGaT has strong perpendicular magnetic anisotropy (PMA). As the temperature increases, thermal fluctuations lead to a gradual weakening of PMA. At 300 K, the AHE loop shows a constricted shape, suggesting the presence of non-collinear spin structures in the FGaT flakes such as labyrinth domains (Fig. S6B and C), which is consistent with previous reports (*3, 43, 44*).

**Pre-set field induced exchange bias effect in FCGT/FGaT device**

The pre-set field protocol serves as the primary methodology in this study to establish and investigate the EB effect. As shown in Fig. 1E, the pre-set field protocol involves zero field-cooling to a target temperature, followed by magnetic field ramping to the pre-set field to induce EB, and subsequent AHE hysteresis loop measurement using a maximum field slightly exceeding the coercive field of the ferromagnetic layer. The ferromagnetic positive and negative coercive fields in the AHE hysteresis loop are designated as $H_C^+$ and $H_C^-$, respectively. A robust interfacial exchange coupling can be established by aligning the magnetic moments at the FCGT/FGaT interface with a field strong enough to switch the Néel order of FCGT. Consequently, an EB field, quantitatively defined as $H_{EB} = (H_C^+ + H_C^-)/2$, is experimentally observed.

Fig. 2A and B display the AHE hysteresis loops measured under distinct temperatures using the pre-set field protocol with ±1 T fields. Notably, the pronounced asymmetry of AHE loops relative to zero field across all temperatures confirms the emergence of EB. The device manifests consistent negative EB (i.e., antiparallel alignment between the pre-set field direction and $H_{EB}$ orientation) over the 5 to 300 K range. The shape of each AHE hysteresis loop exhibits distinct asymmetric characteristic under both positive and negative pre-set magnetic field protocols. The possible reason is that bulk magnetic domains in FGaT demonstrate different response to external field compared with the pinned surface domains (*17*). Notably, magnetization reversal opposing the pre-set field direction displays a gradual multi-step transition, attributed to the sequential overcoming of interfacial exchange coupling. Temperature-dependent $H_{EB}$ and $H_C$ values, calculated from these measurements, are plotted in Fig. 2C and 2D. Here, the coercive field ($H_C$) is defined as $H_C = (H_C^+ - H_C^-)/2$. The $H_{EB}$-temperature relationship exhibits a monotonic decay profile, reaching maximum amplitude (~60 mT) at 5 K and vanishing completely at 300 K, which corresponds to $T_B$ of EB. The asymmetric temperature dependence of $H_{EB}$ under opposite pre-set fields originates from an initially asymmetric domain structure in FGaT prior to zero-field cooling, which may lead to uneven pinning effects from FCGT at the interface. A distinct inflection point emerges at 270 K in both $H_{EB}$-$T$ and $H_C$-$T$ curves, coinciding with the superparamagnetic $T_B'$ derived from the $M$-$T$ curve of FCGT crystal. The broad peak of superparamagnetic transition leads to a larger $T_B$ of EB than the $T_B'$ of superparamagnetism. This temperature also marks the disappearance of spin-flop transitions, suggesting drastically decreased magnetic anisotropy in FCGT above 270 K, which weakens its capacity to pin FGaT magnetic moments. We note that the observed EB effect at RT is not related to the oxidation of FGaT. First of all, the reported $T_B$ is not greater than 140 K, according to a wide range of FGaT-based heterostructures with the natural oxidization studied previously (*24, 27*). Secondly, the observed oxidation layer here is too thin (~ 3 nm) to pin the magnetic moment of the FGaT layer at RT (see Fig. S3D).

To further understand the mechanism of the observed pre-set field-induced EB effect, we performed pulsed magnetic field cycling protocol measurements on Device#1 after ZFC to 5 K. Taking the +1 T pulsed field as an example, each pulse cycle consisted of two sequential operations: 1) application of a +1 T field to induce EB, followed by 2) five consecutive measurements of the AHE hysteresis loops using a maximum sweep magnetic field slightly exceeding the coercive field of the ferromagnetic layer, as shown in Fig. 3A and B. The EB is observed in the first hysteresis loops, and disappeared in the subsequent loops. By further alternating +1 T and -1 T pulsed fields through ten complete cycles, we obtained a total of 50 AHE loops, with the measuring protocol and calculated EB fields $H_{EB}$, as shown in Fig. 3C. The vanishing of the EB effect beyond the second hysteresis loop indicates a strong training effect. The asymmetric behavior of EB and training effect with respect to two different directions of the pre-set field are observed. This might be asymetric domain structure in the FGaT at RT before FC, which induces an aysmmetric domain structure in FCGT after FC. A 1 T pre-set field is insufficient to enforce an collinear alignment between the FGaT and FCGT spins (*45*), resulting in unstable EB and persistent asymmetry under opposite pre-set field polarities.

Subsequently, we implemented an enhanced protocol employing alternating ±6 T pulsed pre-set fields to test the EB effect through ten cycles while maintaining identical measurement parameters (Fig. 3D and E). The result shows distinct behavior compared to that of ±1 T pre-set field pulse. As demonstrated in Fig. 3F, the ±6 T pre-set field induces substantially stronger EB fields ranging 100-150 mT. Remarkably, these enhanced $H_{EB}$ values maintain remarkable stability during subsequent hysteresis loop measurements, demonstrating the suppression of training effect. The sign of EB field can be efficiently reversed upon switching the polarity of pre-set field, demonstrating the reconfigurability of the interfacial magnetic state and EB. This robust EB effect indicates the strong interfacial exchange coupling established by applying pre-set field beyond the spin-flip transition. Furthermore, the AHE measurement shows not only horizontal shifts along *H*-axis but also pronounced vertical offsets, which is related to the uncompensated magnetic moment from the FCGT flake as discussed below. Identical measurements at RT (Fig. S7) reveal preserved EB sign/magnitude controllability by pre-set field but show markedly stronger training effects than at low temperatures, indicating less stable pre-set-field-induced FCGT domain states under thermal conditions.

To understand the origin of the field-tunable EB phenomenon in FCGT/FGaT heterostructures, we employed the macrospin model as the previous reports (*34, 46-48*). In our work, an AFM/FM heterostructure linear-chain model comprising three FCGT layers and one FGaT layer was built to study the pre-set field induced EB effect. As illustrated in Fig. 3G, each vdW layer is represented by a macrospin approximation: the FGaT layer spin corresponds to $FM_1$, while the three layer FCGT spins are designated as $AF_1$, $AF_2$, and $AF_3$ in the order of increasing distance from the FGaT interface. The system's Hamiltonian incorporates three primary energy contributions: Zeeman energy, magnetic anisotropy energy, and exchange interactions.

Both FCGT and FGaT layers exhibit uniaxial magnetic anisotropy. The exchange coupling configuration features ferromagnetic interaction between $FM_1$ (FGaT) and $AF_1$ (nearest FCGT), with antiferromagnetic coupling between successive FCGT layers ($AF_1$-$AF_2$ and $AF_2$-$AF_3$). Detailed mathematical formulations are provided in the Materials and Methods section. The angular evolution of each macrospin under an OOP magnetic field is summarized in Fig. S8, which describes the magnetic field-induced phase transition and pre-set field induced EB. Fig. 3H presents the simulated OOP magnetization evolution under applied fields. The stepped magnetization curve reveals distinct magnetic states, with the central hysteresis loop exhibiting two sharp transitions: 1) the magnetic order of FM layer reverses at critical fields $H_{C1}+$ and $H_{C1}-$, and 2) the uncompensated layer of AFM flips at $H_{C2}+$ and $H_{C2}-$, showing excellent agreement with experimental observations (Fig. S9). The pre-set field lower than $H_{C2}$ would not able to establish interfacial exchange coupling. On the contrary, the high pre-set field leads a switching of Néel vector, further aligning interfacial magnetic moments of FCGT parallel to FGaT. The Néel vector switching is induced by a critical pre-set field between FCGT's spin-flop and spin-flip fields, as shown in Fig. S8A. Subsequent field sweeps below $H_{C2}+$ maintain AFM spin configurations but overcome the AFM-FM exchange coupling, thereby generating measurable EB effect. Furthermore, different direction of pre-set field induces the change in orientation of the Néel order, manifested as vertical offsets in Fig. 3I. This computational prediction precisely matches the experimental phenomena shown in Fig. 3D and E. These combined theoretical and experimental findings conclusively demonstrate that the strong EB effect in FCGT/FGaT systems primarily stems from interfacial exchange coupling.

**Pre-set field induced exchange bias in non-local FCGT/FGaT device**

A non-local device (Device#2) is further fabricated as shown in the inset of Fig. 4A, where FCGT (107 nm) layer partially covers the FGaT (20.2 nm) layer. Channel 1 (CH1) probes the AHE signal from the FCGT/FGaT stacked region, while Channel 2 (CH2) measures the AHE signal from the uncovered FGaT region. Such non-local device enables simultaneous detection of the FGaT region and the FCGT/FGaT heterostacked region under identical conditions. Initially, the AHE hysteresis loops of the device were measured at 200 K under a sweeping magnetic field of 6 T (Fig. 4A). The difference in Hall resistance between distinct magnetization direction, defined as $R_{AHE, FM}$, is smaller in CH1 than that of CH2. This suppression arises from the shunting effect caused by the lower resistivity of the thicker FCGT layer. Notably, CH1 displays a spin-flip transition at 4.2 T due to the presence of FCGT. A weak spin-flip transition is also observed in CH2 under identical magnetic fields, suggesting persistent AFM-FM coupling between FCGT and FGaT even in the non-local configuration, consistent with theoretical calculations (Fig. S8). Subsequently, EB was induced in the non-local device at 100 K via the ±6 T pre-set field (Fig. 4B and C). CH1 and CH2 exhibit identical $H_{EB}$ values, indicating the inherent influence of the FGaT, where the scale of transverse magnetic domain propagation can extend to tens of micrometers (*49*). Such non-local EB effect can be also efficiently switched similar to the local devices. Additionally, the AHE loops with +6 T and -6 T pre-set field demonstrates a lateral shift of 8 mΩ along the $R_{xy}$-axis. This asymmetric behavior again aligns with computational results in Fig. 3I. Another FGaT (29 nm) Hall device (Device#3) was measured by +6 T pre-set field protocol, and no obvious EB is observed at the temperature range from 100 K to 300 K (Fig. 4D), which proves that FCGT is critical to induce EB in our vdW heterostructure.

**Discussion**

In conclusion, we have discovered an perpendicular EB effect in the vdW magnetic heterostructure FCGT/FGaT, with a measured EB field reaching 152 mT (at 5 K under -6 T pre-set field) and a $T_B$ extending up to 300 K. The sign and magnitude of the EB field can be efficiently controlled by manipulating the Néel order of FCGT with magnetic field. Our study reveals that the switching of Néel vector serves as the critical method to control the EB effect. This work not only highlights the significant potential of 2D vdW magnetic heterostructures for investigating interfacial interactions and their manipulation, but also provides insights for developing novel spintronic devices based on interface-engineered low dimensional magnetic systems.

## Materials and Methods

### Single Crystal Synthesis

Single crystals of $(Fe_{0.56}Co_{0.44})_5GeTe_2$ were prepared by the chemical vapor transport (CVT) method. High-purity elemental powders of Fe (99.9%, Aladdin), Co (99.99%, Aladdin), Ge (99.99%, Aladdin), and Te (99.99%, Aladdin) were mixed uniformly in a mortar at a molar ratio of 2.8: 2.2: 1: 2 and then loaded into a quartz tube. Iodine was used as the transport agent. After evacuating and sealing, the quartz tube was placed in a dual-zone tube furnace, where both zones were maintained at 750°C for 10 days, followed by air-cooling to ambient temperature. High-quality single crystals were successfully obtained from the reaction product.

$Fe_3GaTe_2$ single crystals were grown by the self-flux method. High-purity Fe powder (99%, Aladdin), Ga pieces (99.999%, Aladdin), and Te powder (99.99%, Aladdin) were placed in a vacuum quartz tube at a molar ratio of 1: 1: 2 and sealed. First, the mixture was heated to 1000 °C and maintained for 24 h. Then, the temperature was rapidly decrease to 880 °C within 1 hour and then slowly cooled to 780°C within 100 hours. $Fe_3GaTe_2$ single crystals were selected from the ingot. The phase characterization of crystals was analyzed using X-ray diffraction (XRD) with Cu Kα radiation ($\lambda = 1.54056$ Å).

### Magnetic and transport measurements

The magnetism of the $(Fe_{0.56}Co_{0.44})_5GeTe_2$ single crystal and the $Fe_3GaTe_2$ bulk was first characterized by a magnetic property measurement system (MPMS-3; Quantum Design). The FCGT/FGaT heterostructure was prepared by the dry transfer method. The standard Hall bar was patterned by a maskless ultraviolet lithography machine (UV Litho - ACA; TuoTuo Technology) and then etched by a plasma etching machine (IBE-200; ADVANCEDMEMS). The electrodes of the Hall bar were patterned by the maskless ultraviolet lithography machine and then plated with Pt (~50 nm) by magnetron sputtering (ORION 8; AJA International, Inc.). The thickness of the device was measured by an atomic force microscope (Dimension Icon; Bruker). All transport tests were conducted using a physical property measurement system (PPMS DynaCool; Quantum Design).

**Macrospin model simulation**

In the FCGT/FGaT heterostructure linear-chain model, the intralayer coupling is much stronger than the interlayer coupling. The magnetization intensity of each layer can be regarded as a macrospin and is coupled with the nearest neighbor layer through the interlayer exchange interaction. Assuming that the entire FGaT layer can be regarded as one macroscopic spin, the Hamiltonian of the N-layer FCGT/FGaT system can be expressed as:

$$\widehat{H} = -\sum_{n=1}^{N} H \cdot M_n \cos\theta_n + \sum_{n=1}^{N} K_n (\sin\theta_n)^2 + \sum_{n=1}^{N-1} J_n \cos(\theta_{n+1} - \theta_n)$$

Among them, $H$ represents the magnetic field applied parallel to the $c$ axis, $M_n$, $K_n$ and $\theta_n$ denote the saturation magnetization, magnetic anisotropy energy of the $n$th spin and is the angle between the macrospin of the $n$th spin and the $c$ axis, respectively. $J_n$ stands for the exchange interaction between $n$th and $(n+1)$th spins. By incorporating parameters: $M_1 = 15$, $M_n = 1$ (n > 1), $K_1 = 3$, $K_n = 0.5$ (n > 1), $J_1 = 1$, $J_n = -3$ (n > 1) into the linear-chain model of the heterostructure, we solve for the evolution of spins in each layer under external magnetic fields when the system reaches its lowest energy state. This allows us to determine the overall magnetization variation of the heterostructure as a function of the external field. All simulations are performed using MATLAB.

**Acknowledgement**

**Funding:**

National Key R&D Program of China grant no. 2022YFA1203902 (Y.H., R.W.)

National Natural Science Foundation of China (NSFC) grant nos. 12241401 (J.Y.), 12374108 (R.W.), 12104052 (R.W.), 52471249 (L.P.), 52027801(Y.H.), and 92263203(Y.H.)

Guangdong Provincial Quantum Science Strategic Initiative grant no. GDZX2401002 (R.W., F.Z.))

GJYC program of Guangzhou grant no. 2024D01J0087 (R.W.)

Fundamental Research Funds for the Central Universities (R.W.)

**Author contributions:**

Conceptualization: RW, SD, ZL

Single crystal growth: JW, BD, ZH

Device preparation and transport measurements: JW

Macrospin modelling: RW

Transmission Electron Microscopy characterization: LP, YJ, WS, FZ

Data analysis and interpretation: RW, SD, JY, YY

Supervision and financial support: YH and RW

Writing—original draft: JW

Writing—review & editing: all authors

**Figures and Tables**

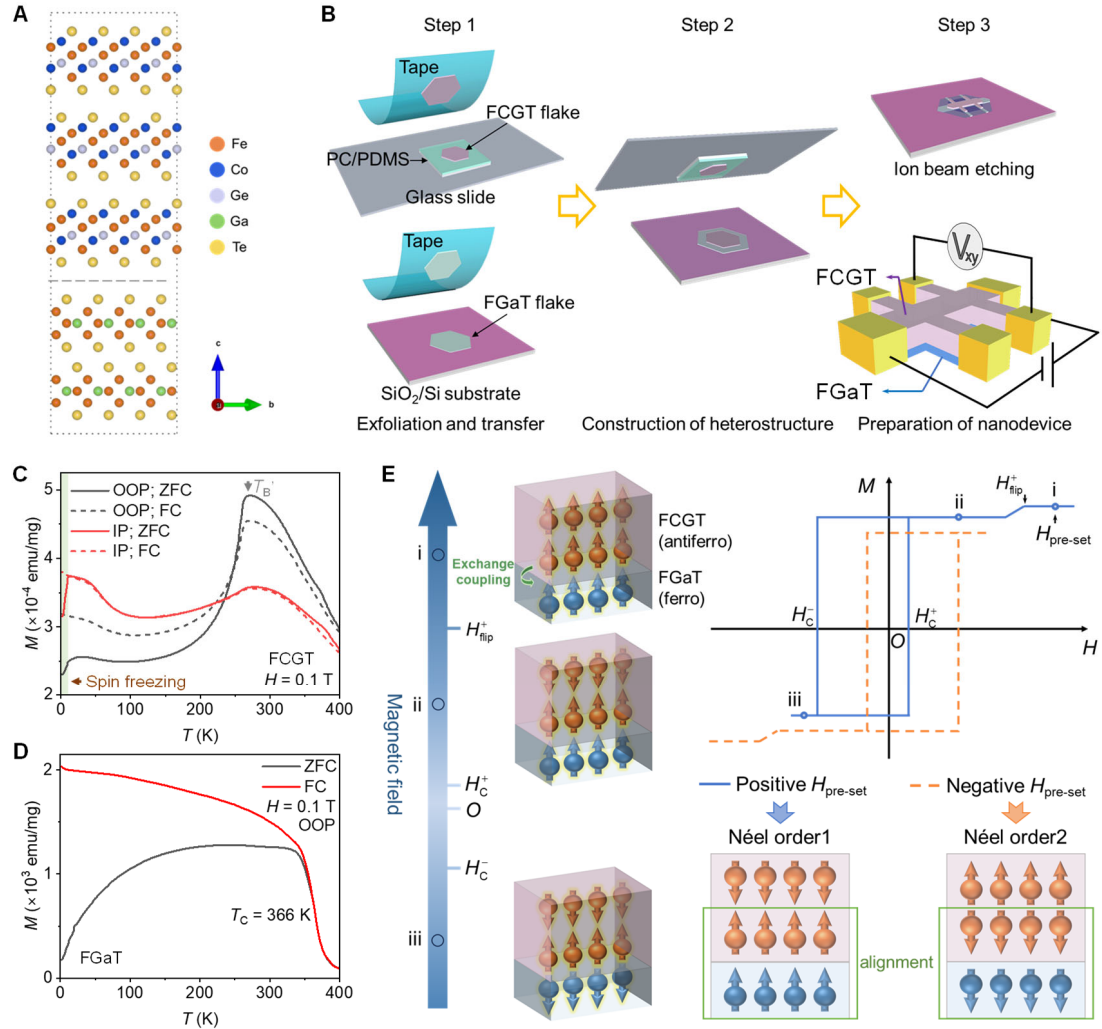

**Fig. 1. Fabrication of (Fe$_{0.56}$Co$_{0.44}$)$_5$GeTe$_2$ (FCGT)/Fe$_3$GaTe$_2$ (FGaT) heterostructure device and pre-set field protocol description.** (**A**) Crystal structure diagram of the FCGT/FGaT heterostructure with crystallographic *a*-axis alignment. (**B**) Schematic illustration of the fabrication process for FCGT/FGaT heterostructure and Hall bar device. (**C**) Temperature-dependent magnetization curves (*M-T* curves) of FCGT crystal with a fixed applied field, obtained under $H = 0.1$ T during zero field-cooling (ZFC) and field-cooling (FC) in two magnetic field directions, out-of-plane (OOP) and in-plane (IP). (**D**) Temperature-dependent magnetization curves of FGaT bulk under $H = 0.1$ T during ZFC and FC. (**E**) Schematic diagram of the pre-set field induced exchange bias (EB) effect.

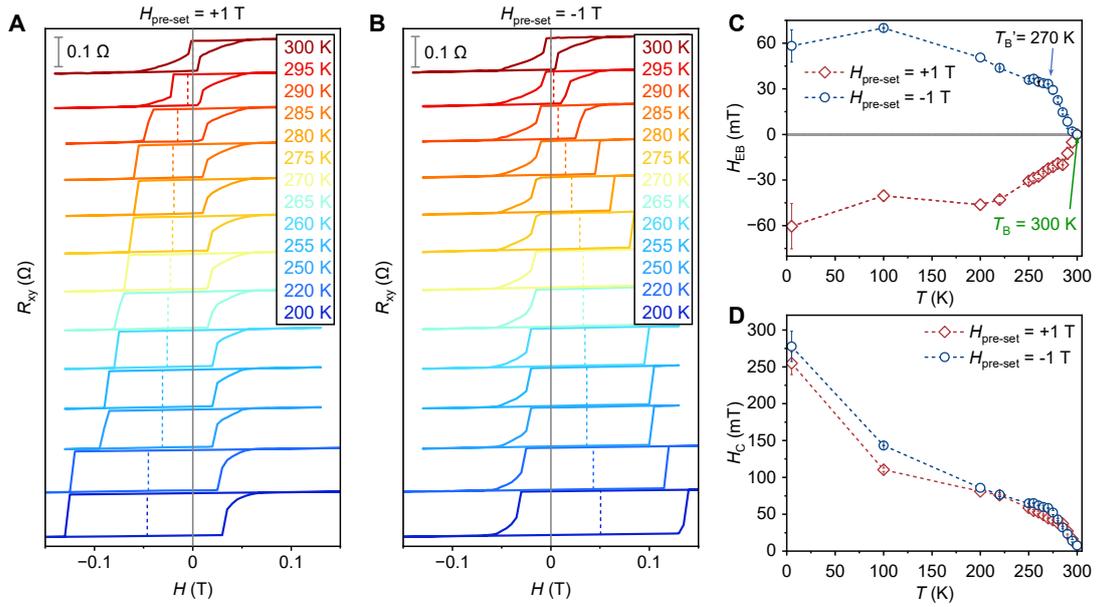

**Fig. 2. Emergence of exchange bias (EB) phenomena in FCGT/FGaT heterostructure device.** The anomalous Hall effect (AHE) hysteresis loops measured at different temperatures for FCGT/FGaT Hall bar device (Device#1) under (**A**) +1 T and (**B**) -1 T pre-set field protocols, respectively. Summarized temperature-dependent curves of (**C**) EB field $H_{EB}$ and (**D**) coercive field $H_C$ under +1 T pre-set field and -1 T pre-set field protocols, respectively. Three loops were measured at each temperature to obtain mean values and standard deviations.

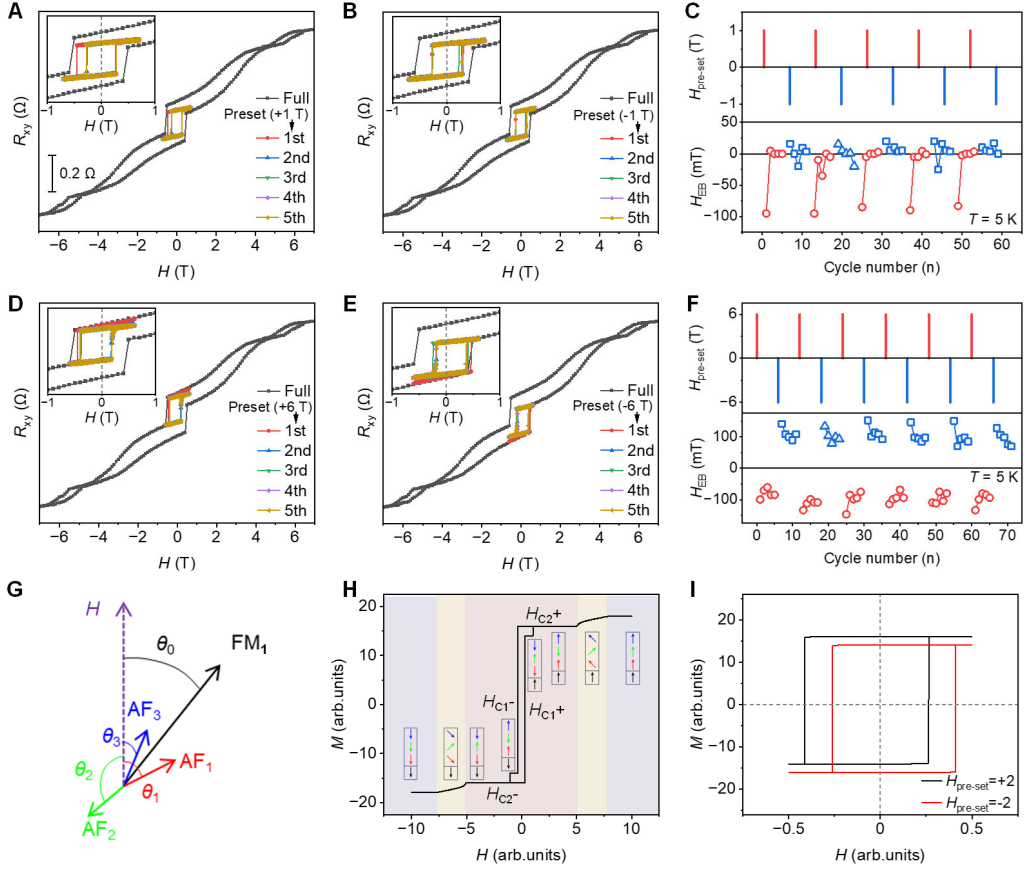

**Fig. 3. Harnessing strong exchange coupling via field alignment of Néel order in FCGT/FGaT device.** Pulsed magnetic field cycling protocol measurements on Device#1. All measurements were performed at 5 K with an OOP magnetic field orientation. (**A**) and (**B**) AHE hysteresis loops of FCGT/FGaT Device#1 after pulsed +1 T and −1 T cycling protocol, respectively, subjected to five cycling tests, compared with full AHE signals measured under a 7 T sweep magnetic field. (**C**) Schematic of the pulsed ±1 T cycling protocol and the calculated hysteresis-based $H_{EB}$ as a function of cycle number n. (**D**) and (**E**) AHE hysteresis loops of FCGT/FGaT Device#1 pulsed +6 T and −6 T cycling protocol, respectively, subjected to five cycling tests, compared with full AHE signals. (**F**) Schematic of the pulsed ±6 T cycling protocol and the calculated hysteresis-based $H_{EB}$ as a function of cycle number n. (**G**) Schematic of the macrospin model for the AFM/FM heterostructure (composed of three FCGT layers and one FGaT layer). (**H**) Magnetic field-dependent magnetization curve of AFM/FM macrospin model. (**I**) Hysteresis loops of AFM/FM macrospin model simulated under different pre-set field protocols, revealing not only a significant EB effect but also a distinct vertical shift along the *M*-axis.

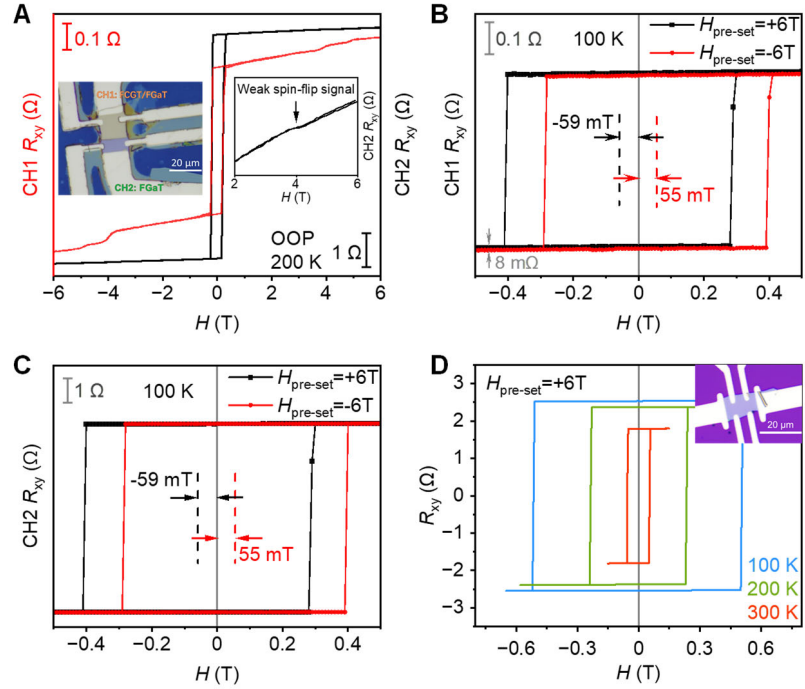

**Fig. 4. EB in non-local FCGT/FGaT device.** (**A**) OOP AHE hysteresis loops measured at 200 K for non-local FCGT/FGaT Device#2 under 6 T sweep field to probe AFM phase transition characteristics. Left inset displays the optical image of Device#2, while the right inset shows detailed AHE signals from channel 2 (CH2). AHE hysteresis loops recorded at 100 K for (**B**) channel 1 (CH1) and (**C**) CH2 of Device#2 under ±6 T pre-set field protocols, respectively. (**D**) AHE hysteresis loops measured at different temperatures for FGaT Hall bar device (Device#3) under +6 T pre-set field protocol. The inset displays the optical image of Device#3. No distinct EB can be observed.